\documentclass[aps,pra,twocolumn,floatfix]{revtex4-2}

\usepackage{natbib,color}
\usepackage[utf8]{inputenc}

\usepackage{graphicx}
\usepackage{amsmath}
\usepackage{amssymb}

\newcommand{\tcr}[1]{\textcolor{black}{#1}}

\begin{document}

\title{Fundamental limits on determination of photon number statistics from measurements with multiplexed on/off detectors}

\author{Jarom\'{i}r Fiur\'{a}\v{s}ek}
\email{fiurasek@optics.upol.cz}
\affiliation{Department of Optics, {Faculty of Science}, Palack\'y University, 17. listopadu 12, 77900 Olomouc, Czech Republic}

\begin{abstract}
We investigate fundamental bounds on the ability to determine photon number distribution and other related quantities from tomographically incomplete measurements with an array of $M$ detectors that can only distinguish the absence or presence of photons. We show that the lower and upper bounds on photon number probabilities can be determined by solving a linear program.  We present and discuss numerical results for various input states including thermal states, coherent states, squeezed states, and highly non-classical single-photon subtracted squeezed vacuum states. Besides photon number probabilities we also investigate bounds on the parity of photon number distribution that determines the value of Wigner function of the state at the origin of phase space. Moreover, we also discuss estimation of mean photon number as an example of a quantity described by an unbounded operator. Our approach and results can provide quantitative guidance on the number of detection channels required to determine the photon number distribution with a given precision.
\end{abstract}

\maketitle

\section{Introduction}

Experimental determination of photon number distribution is a long studied problem in quantum optics. Besides being of fundamental interest, the detection and counting of photons represents an important ingredient of many protocols and schemes in optical quantum information processing and quantum metrology. As the technology progresses, detectors that can directly count the number of photons have become available \cite{Lita2008,Kim1999,Gerrits2012,Harder2016,Sperling2017,Cahall2017,Magana2019,Liao2020,Zhu2020,Eaton2022,Davis2022,Los2024}. Nevertheless, many commonly employed detectors can only distinguish the presence and absence of photons. Approximate photon-number resolving detectors can be constructed from such binary on/off detectors by spatial or temporal multiplexing \cite{Paul1996}. The signal is split into $M$ modes and sent onto an array of $M$ detectors, and the number of detector clicks provides information about the photon number distribution. Such multiplexed photon detectors have been thoroughly investigated  and utilized in numerous experiments \cite{Rehacek2003,Banaszek2003,Fitch2003,Achilles2003,Bartley2013,Kalashnikov2011,Sperling2012a,Sperling2012b,Mattioli2016,Kroger2017,Sperling2017b,
Zhu2018,Straka2018,Kovalenko2018,Jonsson2019,Tiedau2019,Lachman2019,Hlousek2019,Cheng2022,Knoll2023,Hlousek2024,Sullivan2024,Krishnaswamy2024,Santana2024,Banner2024}.

Since the number of detection channels $M$ is  finite, the click statistics measured by the multiplexed detector generally does not fully and uniquely specify the photon number distribution.
In other words, infinitely many different photon number distributions can yield the same  cilick statistics.  One can attempt to remove this ambiguity by making some additional assumptions, e.g. that the photon number distribution has the maximum entropy compatible with the observed click statistics \cite{Hlousek2019}. However, such additional assumptions may not be generally justified, especially if the goal is to characterize artificially generated quantum states with non-trivial photon number distributions.

 In the present work, we avoid such additional assumptions and we investigate the resulting  fundamental uncertainty of determination of the photon number distribution from the measured click statistics. The click statistics allows us to obtain lower and upper bounds on the photon number probabilities $p_n$ as well as on other quantities that can be expressed as linear combinations of $p_{n}$.  We show that these lower and upper bounds can be determined by solving a suitably formulated linear program. We investigate the behavior of these bounds for photon number probabilities $p_n$ and photon number parity operator, which specifies the value of the Wigner function of the state at the origin of phase space. 

 For a given fixed photon number distribution the estimation precision increases with the increasing number of detection channels $M$ and also increases with increasing detection efficiency $\eta$. Such behavior is certainly not surprising, but our approach allows us to precisely quantify the impact of $M$ and $\eta$ on the uncertainty of determination of $p_n$. As each detector and detection channel represents a valuable resource, our analysis can be used to identify the minimum number of detection channels $M$ to achieve the required measurement and estimation precision. 

We also investigate the determination  of the mean photon number $\bar{n}$. Since the photon number operator is unbounded, the click statistics generally does not  yield a finite upper bound on $\bar{n}$. However, we show that by making some minimal assumptions about the maximum photon number that can non-negligibly contribute to $\bar{n}$, useful estimates of $\bar{n}$ can be obtained even for moderate number of detection channels $M$.  
We complement our numerical analysis by examples of simple analytical bounds on the single photon probability $p_1$ and the two-mode probability $p_{1,1}$ that a single photon is present in each mode. 

The rest of the paper is organized as follows. In Sec.~II we describe the considered measurement scheme and formulate the linear program that is used to obtain the fundamental 
bounds on photon number probabilities $p_{n}$ and other related quantities. Numerical results are presented and discussed in Sec.~III. Determination of mean photon number from click statistics is analyzed within our framework in Sec.~IV. Examples of simple analytical bounds on probabilities $p_1$ and $p_{1,1}$ are given in Sec.~V. Finally, Sec.~VI contains a brief discussion and conclusions.

\begin{figure}[t]
\includegraphics[width=0.85\linewidth]{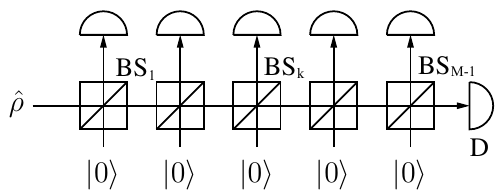}
\caption{Multiplexed detector of photons. The input signal is evenly split among $M$ output modes, e.g., by an array of \tcr{$M-1$ unbalanced} beam splitters \tcr{BS$_k$} with suitably chosen transmittances \tcr{$T_k=(M-k)/(M-k+1)$}.  Each output mode is measured with a binary detector $D$ that can distinguish the presence and absence of photons. The number of detector clicks $m$ represents the measured signal. }
\end{figure}

\section{Measurement scheme}
The considered measurement scheme is depicted in Fig. 1. The measured optical beam is evenly split into $M$ channels and each channel is measured with a binary detector that can distinguish the presence and absence of photons. \tcr{The scheme in Fig.~1 utilizes a sequence of unbalanced beam splitters but also other configurations are possible, e.g. a cascaded tree-like structure where each beam is repeatedly split into two by a balanced beam splitter. To achieve precise balancing of the scheme, tunable beam splitters are required, which can be implemented for instance with the use of wave plates and polarizing beam splitters \cite{Hlousek2019}.}
The measured click statistics $c_m$ represent probabilities that exactly $m$ detectors click simultaneously. Assuming total detection efficiency $\eta$ to be the same for each detector, the click statistics is determined by the photon number distribution $p_n$ of the measured optical beam as follows,
\begin{equation}
c_m=\sum_{n=m}^\infty C_{mn} p_n,
\label{cmdefinition}
 \end{equation}
where \cite{Hlousek2019}
\begin{equation}
C_{mn}= {M \choose m} \sum_{j=0}^m (-1)^j {m \choose j} \left[1-\eta+ \frac{(m-j)\eta}{M}\right]^n.
\end{equation}
Sampling of the click statistics (\ref{cmdefinition}) is equivalent to measurement of probabilities of projection onto vacuum $q_{0,k}$ of the input state transmitted through  lossy  channels with transmittances $T_k=\eta k/M$, $0 \leq k \leq M$ \cite{Mogilevtsev1998}
\begin{equation}
q_{0,k}=\sum_{n=0}^\infty (1-T_k)^n p_n.
\label{q0kdefinition}
\end{equation}
It holds that
\begin{equation}
c_{m}={M \choose m} \sum_{j=0}^m (-1)^j {m \choose j} q_{0,M-m+j}.
\label{cmvacuumprobability}
\end{equation}
This implies that instead of the spatially multiplexed scheme in Fig. 1 \tcr{one} could also in principle utilize a scheme that includes only a single on/off detector and the transmittance $T_k$ is set by  tunable attenuator \cite{Mogilevtsev1998,Rossi2004}. By performing measurements for various $T_k$, the probabilities $q_{0,k}$  can be sampled and the click statistics $c_m$ recovered from Eq. (\ref{cmvacuumprobability}).

\begin{figure}[t]
\includegraphics[width=\linewidth]{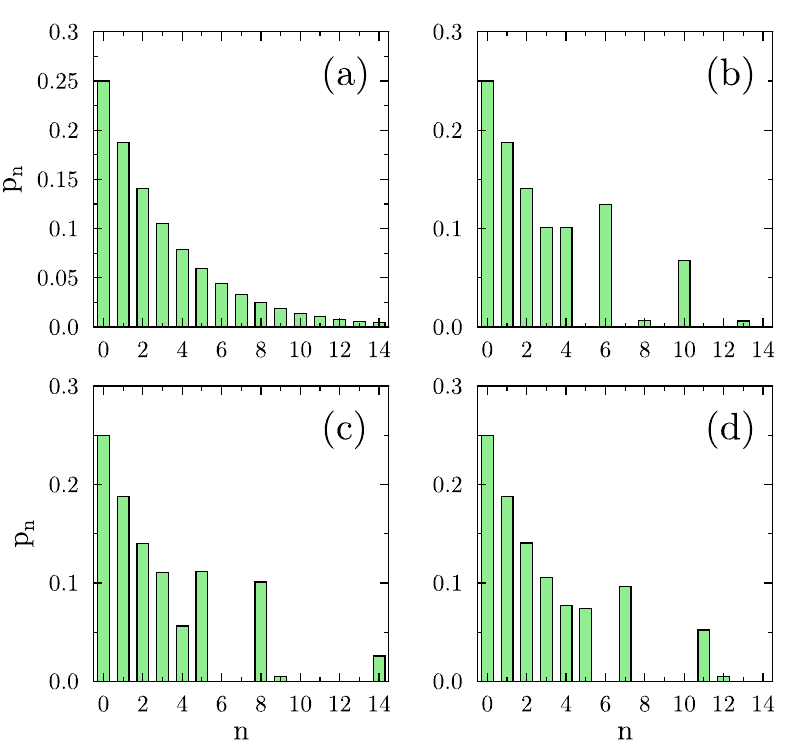}
\caption{\tcr{Four different photon number distributions $p_n$ are displayed that all yield the same click statistics when measured with balanced multiplexed detector with $M=10$ channels and $\eta=1$. The click statistics corresponds to statistics generated by thermal state with mean photon number $\bar{n}=3$. For clarity, only $p_n$ for $n \leq 14 $ are shown in the figure.}}
\label{figfourdistributions}
\end{figure}

Measurement with the multiplexed detector yields $M$ independent parameters. This means that the measurement is tomographically incomplete and the probabilities $p_n$ cannot be fully unambiguously reconstructed from the click statistics $c_m$. Many different distributions $p_n$ can yield exactly the same click statistics $c_{m}$ and are therefore fundamentally indistinguishable by the measurement. \tcr{This is illustrated in Fig.~\ref{figfourdistributions}, where we plot four different probability distributions that yield exactly the same click statistics when measured with ten-port balanced detector. Based solely on the click statistics it is thus fundamentally impossible to distinguish these photon number distributions. In Ref. \cite{Hlousek2019}, it was proposed to resolve this ambiguity by essentially selecting the distribution that possesses maximum entropy, which was technically accomplised by the so-called expectation-maximization-entropy (EME) algorithm. However, the selection of just a single probability distribution may not provide an accurate picture.  For example, characterization of a source of  thermal states by EME approach may yield overly optimistic results, as the Bose-Einstein distribution plotted in Fig.~\ref{figfourdistributions}(a) exhibits the highest entropy among the four distributions plotted in Fig.~\ref{figfourdistributions}. 
For tomographically inclomplete measurements, interval estimates or posterior distributions of $p_n$ can provide more accurate representation of the information contained in the click statistics.} 

\tcr{Specifically, the knowledge of click statistics restricts} the region of allowed $p_n$. For each $p_n$ it \tcr{is} possible to determine a region of allowed values $p_n \in [p_{n,\mathrm{min}},p_{n,\mathrm{max}}]$. Similar \tcr{intervals of possible values} can also be establised  for physical quantities $Z$ that are linear functions of $p_n$, 
\begin{equation}
Z=\sum_{n=0}^\infty z_n p_n,
\label{Zdefinition}
\end{equation}
provided that the coefficients $z_n$ are bounded and there exists finite $B>0$ such that $|z_n|<B$ holds for all $n$.

In this work we are interested in the effect of finite number of detection channels. We will therefore assume that the true values of $c_m$ (or, equivalently, $q_{0,k}$) are known. In most cases we shall assume $\eta=1$ but the presented methods can be straightforwardly extended to imperfect detectors with $\eta<1$. We provide below explicit examples illustrating the effect of 
$\eta$  on determination of $p_n$.

In order to determine the maximum and minimum values of $Z$ compatible with the observed click statistics, we should maximize or minimize $Z$ under the equality constraints given by Eq. (\ref{cmdefinition}) or equivalently by Eq. (\ref{q0kdefinition}), and the additional positivity  constraints 
\begin{equation}
p_n \geq 0.
\end{equation}
This is an instance of a linear program. We can write all the equality constraints in general form as 
\begin{equation}
b_k =\sum_{n=0}^\infty A_{kn} p_n. 
\label{equalityconstraints}
\end{equation}
\tcr{Specifically, for the balanced $M$-channel multiplexed detector we can either set $A_{kn}=C_{kn}$ and $b_k=c_k$ or equivalently we can set $A_{kn}=(1-\eta k/M)^n$ and $b_k=q_{0,k}$. Note, however, that the formulation (\ref{equalityconstraints}) is general and specific choices of $A_{kn}$ can describe a wide range of detectors. 
Note also that the probability normalization constraint
$\sum_{n=0}^\infty p_n=1$
is automatically incorporated in Eqs. (\ref{cmdefinition}) and (\ref{q0kdefinition}) because 
\begin{equation}
\sum_{m=0}^M c_{m}=1, \qquad \sum_{m=0}^M C_{mn}=1,
\end{equation}
and $q_{0,0}=\sum_{n=0}^{\infty} p_n=1$ holds by definition.}

 Since $n$  is unbounded, we must truncate the photon number distribution at some finite cutoff $N$ when performing the calculations. Assuming that we know the true photon number distribution $p_{n,\mathrm{true}}$ we can  formulate a linear program for a finite vector $\vec{x}=(x_0,\ldots,x_N)$:
\begin{equation}
\begin{array}{c}
\mathrm{minimize} \quad \displaystyle{\tilde{Z}=\sum_{n=0}^N z_n x_n, } \\[4mm]
\mathrm{under~the~constraints} \\[2mm]
 x_n \geq 0,  \quad 0 \leq n \leq N,\\[2mm]
\displaystyle{\tilde{b}_k=\sum_{n=0}^N A_{kn} x_n,} \quad 0 \leq k\leq  M,
\end{array}
\label{linearprogram}
\end{equation}
where 
\begin{equation}
\tilde{b}_k=\sum_{n=0}^N A_{kn} p_{n,\mathrm{true}}.
\end{equation}
The advantage of this approach is that the linear program (\ref{linearprogram}) is feasible by construction, because the vector $x_n=p_{n,\mathrm{true}}$, $0\leq n\leq N$, obviously satisfies all the constraints. 
 In order to account for the tail of the true distribution, the lower and upper bounds corresponding to optimization over $p_{n}$ up to $n=N$ can be calculated as 
\begin{equation}
Z_{\mathrm{min}}=\tilde{Z}_{\mathrm{min}}+\delta Z, \quad Z_{\mathrm{max}}=\tilde{Z}_{\mathrm{max}}+\delta Z,
\end{equation}
where
\begin{equation}
\delta Z=\sum_{n=N+1}^\infty z_n p_{n,\mathrm{true}}.
\end{equation}
Maximum and minimum values of $p_m$  can be obtained by choosing $z_n=\pm \delta_{mn}$ and the tail $\delta Z$ vanishes in this case.
The true lower and upper bounds are obtained in the limit $N \rightarrow \infty$. \tcr{If the observable $Z$ is bounded and $|z_n|\leq B$, then the tail $\delta Z$ can be bounded by the inequality
\[
|\delta Z| \leq B \sum_{n=N+1}^\infty p_{n,\mathrm{true}}=B \Sigma_N.
\]
Since  $\lim_{N\rightarrow \infty }\Sigma_N=0$  holds for any given fixed probability distribution $p_{n,\mathrm{true}}$,  the tail $\delta Z$ will vanish in the asymptotic limit.}
In practice, it is sufficient to choose $N$ large enough such that $\Sigma_N$  becomes negligibly small. Examples of dependence of the lower and upper bounds on $N$ are given below.

Optimality of the solution of the linear program (\ref{linearprogram}) can be verified by solving the corresponding dual program
\begin{equation}
\begin{array}{c}
\mathrm{maximize} \quad \displaystyle{\tilde{Y}=\sum_{k=0}^M y_k \tilde{b}_k, } \\[5mm]
\mathrm{under~the~constraints} \\[2mm]
\displaystyle{\sum_{k=0}^M y_k A_{kn}  \leq z_n, \quad 0\leq n\leq N}.
\end{array}
\label{dualprogram}
\end{equation}
Any feasible solution $\tilde{Y}$ of the dual program (\ref{dualprogram}) provides a lower bound on the solution of the primal program, $\tilde{Z}_{\mathrm{min}} \geq \tilde{Y}$. Strong duality implies that equality holds for the optimal solutions, $\tilde{Y}_{\mathrm{opt}}=\tilde{Z}_{\mathrm{opt}}$. Since $\tilde{Y}_{\mathrm{opt}}$ is also a lower bound on $\tilde{Z}_{\mathrm{opt}}$, this  proves the optimality of the  solution. 

\begin{figure}[t]
\includegraphics[width=\linewidth]{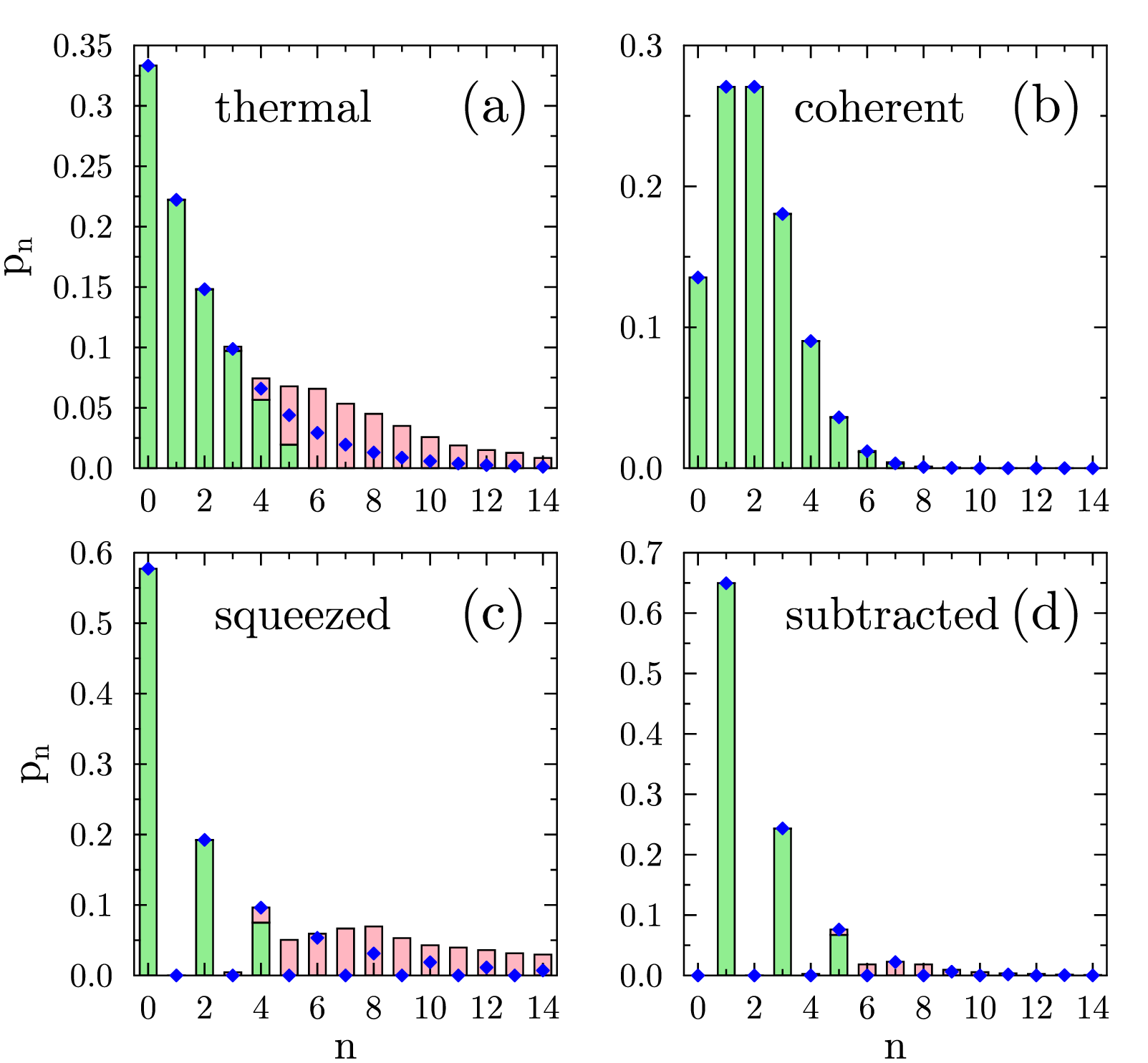}
\caption{Lower bounds (green bars) and upper bounds (pink bars) on the photon number probabilities  $p_{n}$  are plotted for four  different states with $\bar{n}=2$: thermal state (a), coherent state (b), squeezed vacuum state (c), and single-photon subtracted squeezed vacuum state (d). Blue diamonds indicate the true photon number distributions. Parameters of the multiplex detector read $M=10$ and $\eta=1$, and the photon number cutoff  was set  to $N=80$ in the numerical calculations.}
\label{figpnbounds}
\end{figure}

\section{Numerical results}

We have performed numerical calculations for four different types of quantum states: thermal states with Bose-Einstein photon number distribution,
\begin{equation}
p_{n,\mathrm{thermal}}=\frac{1}{\bar{n}+1} \left( \frac{\bar{n}}{\bar{n}+1}\right)^n,
\label{pnthermalstate}
\end{equation}
coherent states with Poisson photon number distribution,
\begin{equation}
p_{n,\mathrm{coh}}=\frac{\bar{n}^n}{n!} e^{-\bar{n}},
\label{pncoherentstate}
\end{equation}
squeezed vacuum states that exhibit oscillations in $p_n$,
\begin{equation}
p_{2n}=\frac{(2n)!}{2^{2n} n!^2}\frac{(\tanh r)^{2n}}{\cosh r}, \qquad p_{2n+1}=0,
\label{pnsqueezedvacuum}
\end{equation}
and the single-photon subtracted squeezed vacuum state,
\begin{equation}
p_{2n-1}=\frac{(2n)!}{2^{2n} n!^2}\frac{2n (\tanh r)^{2n}}{\sinh^2 r \cosh r}, \qquad p_{2n}=0.
\label{pnsubtracted}
\end{equation}
Here $\bar{n}$ denotes the mean photon number and $r$ is the squeezing constant. For the squeezed vacuum state (\ref{pnsqueezedvacuum}) we have $\bar{n}=\sinh^2 r$ and for the single-photon subtracted squeezed vacuum state we obtain $\bar{n}=1+3\sinh^2 r$.

\begin{figure}[t]
\includegraphics[width=\linewidth]{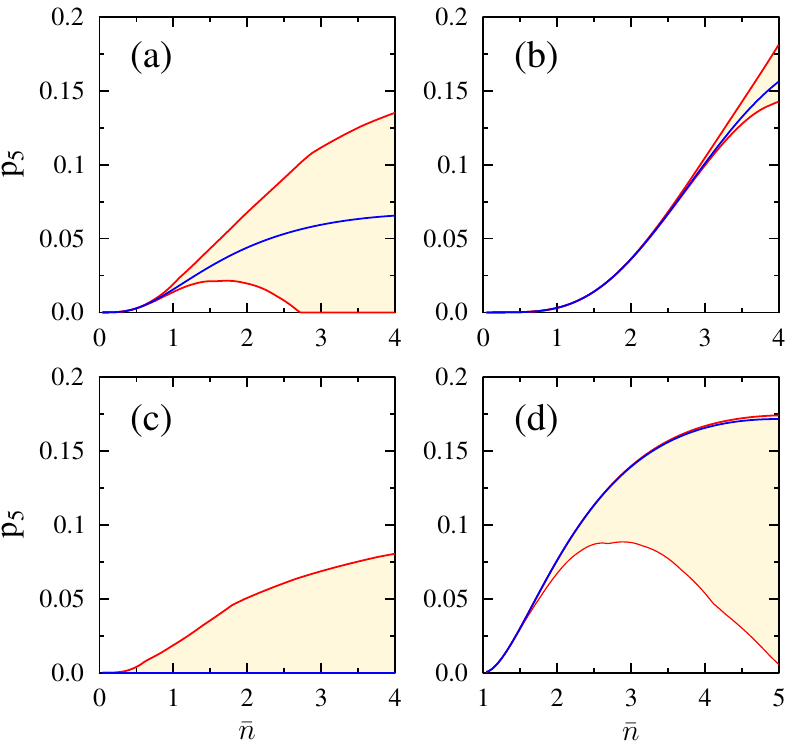}
\caption{Lower and upper bounds on photon number probability $p_5$ are plotted in dependence on the mean photon number $\bar{n}$  for four different states:  thermal state (a), coherent state (b), squeezed vacuum state (c), and single-photon subtracted squeezed vacuum state (d). Blue lines indicate the true values of $p_5$. Parameters of the multiplexed detector read $M=10$ and $\eta=1$. }
\label{figp5bounds}
\end{figure}

The linear program (\ref{linearprogram}) was solved with software Mathematica using the function LinearProgramming. 
In most cases, the, the simplex method was utilized to carry out the optimization, which ensures that exact optimal solution is found that exactly satisfies all the constraints. The simplex algorithm is considerably slower than the interior point methods. Nevertheless, for the studied problems the calculations were fast enough even with the simplex algorithm.

The numerically determined lower and upper bounds on $p_n$ are plotted in Fig.~\ref{figpnbounds} for $\bar{n}=2$ and $M=10$. The cutoff was set to  $N=80$ which is large enough to ensure that the truncation has no effect on the reported results.  We can observe that rather large uncertainties in the determination of $p_n$ occur for the thermal state and squeezed vacuum state. These states exhibit large variances of the photon number operator $\hat{n}$ and for the chosen mean photon number  $\bar{n}$ the population of higher Fock states with $n>M$  is non-negligible, e.g. $p_{11,\mathrm{thermal}}=0.0039$. By contrast, the uncertainties of determination of photon number distribution of the coherent state with $\bar{n}=2$ is very small, because  the population of Fock states above $10$ is negligible, e.g. $p_{11,\mathrm{coh}}=6.94\times 10^{-6}$. The subtraction of a photon from a squeezed vacuum state results in a narrower photon number distribution as compared to the input squeezed vacuum state, which leads to smaller uncertainties in estimation of $p_n$, see Fig.~\ref{figpnbounds}(d).  Note also that the probabilities $p_n$ of lowest photon numbers $n$ are determined with high precision. The uncertainty 
\begin{equation}
\Delta p_n=p_{n,\mathrm{max}}-p_{n,\mathrm{min}}, 
\end{equation}
 first increases with $n$ but then it starts to decrease again because having large populations of high Fock states $|n\rangle$ with $n \gg M$ is incompatible with the constraints imposed by $c_m$ or equivalently by $q_{0,k}$.

\begin{figure}[t]
\includegraphics[width=0.9\linewidth]{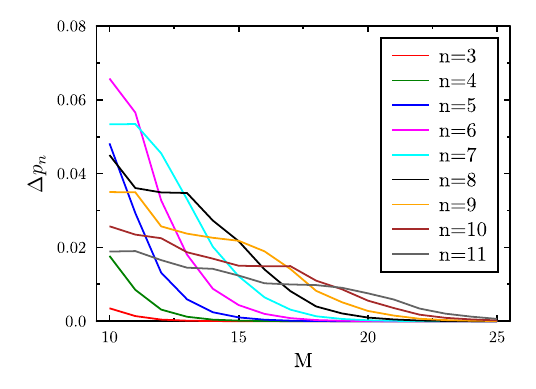}
\caption{The photon number uncertainties $\Delta p_n$ are plotted as functions of  the number of detector channels $M$ for thermal state with $\bar{n}=2$. Photon number cutoff is set to $N=80$ and $\eta=1$ is assumed in the numerical calculations.}
\label{figMdependence}
\end{figure}

To further illustrate the dependence of $\Delta p_n$ on the width of the photon number distribution and the number of the detection channels, we plot in Fig.~\ref{figp5bounds} the dependence of the lower and upper bounds  $p_{5,\mathrm{min}}$  and $p_{5,\mathrm{max}}$  as functions of the mean photon number $\bar{n}$ for the four considered photon number distributions. As expected, for fixed number of detection channels $M$ the uncertainty increases with increasing width of the distribution, here quantified by $\bar{n}$.  Figure~\ref{figMdependence} shows the dependence of the uncertainties $\Delta p_n$  on $M$ for a thermal state with fixed photon number $\bar{n}$. As $M$ increases, more information is gathered about the photon number distribution and the constraints on $p_n$ become tighter. Consequently $\Delta p_n$ decrease with increasing $M$.

\begin{figure}[t]
\includegraphics[width=\linewidth]{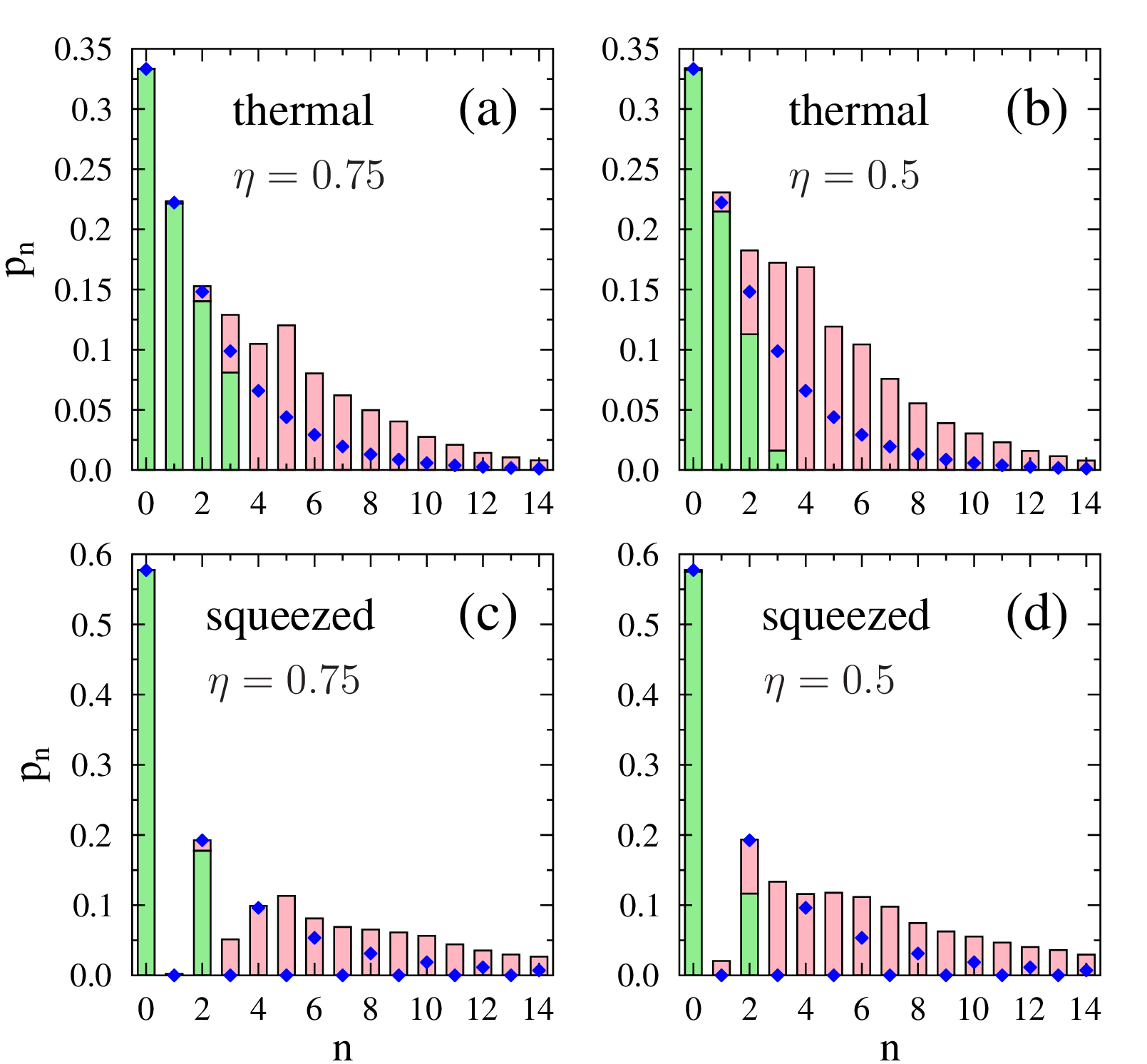}
\caption{Dependence of the uncertainty of determination of $p_{n}$ on detection efficiency $\eta$. The lower bounds (green bars) and upper bounds (pink bars) on the photon number probabilities  $p_{n}$  are plotted for thermal state (a,b) and squeezed vacuum state (c,d) with $\bar{n}=2$. Blue diamonds indicate the true photon number distributions. Parameters of the multiplexed detector read $M=10$, $\eta=0.75$ (a,c), and $\eta=0.5$ (b,d).}
\label{figetadependence}
\end{figure}

The influence of detection efficiency $\eta$ on determination of $p_n$ is illustrated in Fig.~\ref{figetadependence}. We can observe that the uncertainties $\Delta p_{n}$ increase with decreasing detection efficiency, which is the expected behavior. Detection with efficiency $\eta$ is equivalent to measurement with perfect detectors of a state transmitted through a lossy channel with transmittance $\eta$.
The virtual perfect detectors then measure state with modified photon number distribution \cite{Zambra2006}
\begin{equation}
\tilde{p}_n=\sum_{m=n}^\infty {m \choose n} \eta^n (1-\eta)^{m-n} p_m.
\label{pnlosses}
\end{equation}
On the one hand, the distribution $\tilde{p}_n$ has reduced width with respect to the original photon number distribution $p_n$.  For instance, for input thermal or coherent state the losses only reduce the mean photon number of the state to $\eta \bar{n}$. On the other hand, to recover the original distribution $p_m$ from $\tilde{p}_{n}$ one has to invert relation (\ref{pnlosses}). Overall, this results in increased uncertainties $\Delta p_n$ for lossy detection.

\begin{figure}[b]
\includegraphics[width=\linewidth]{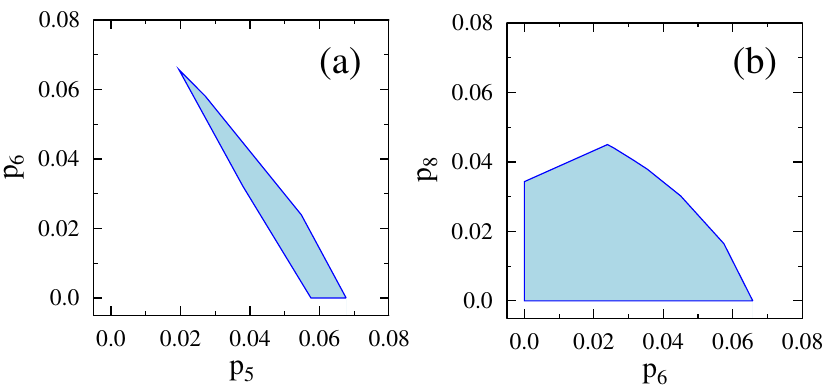}
\caption{\tcr{ Areas of probability pairs $[p_j, p_k]$ compatible with a given click statistics. The results are plotted for click statistics generated by thermal state with $\bar{n}=2$, $M=10$, $\eta=1$. Results are shown for two probability pairs $[p_5,p_6]$ (a) and $[p_6,p_8]$ (b).}}
\label{figpnpairs}
\end{figure}

\tcr{So far, we have plotted the individual lower and upper bounds on $p_n$ obtained by minimization or maximization of $p_n$ under the constraints imposed by the available click statistics. However, these individual bounds cannot be always attained simultaneously due to the presence of linear constraints on $p_n$. To illustrate this, we plot in Fig.~\ref{figpnpairs} the areas of  probability pairs $[p_j,p_k]$ that are compatible with a given click statistics. We can see that the areas are not simple rectangles which indicates the presence of correlations or anticorrelations. The boundaries of the convex sets plotted in Fig.~\ref{figpnpairs} were determined by maximization of $p_j\cos(\phi)+p_k\sin(\phi) $ for $100$ different values of $\phi$ in the interval $[0,2\pi]$.}

\begin{figure}[t]
\includegraphics[width=\linewidth]{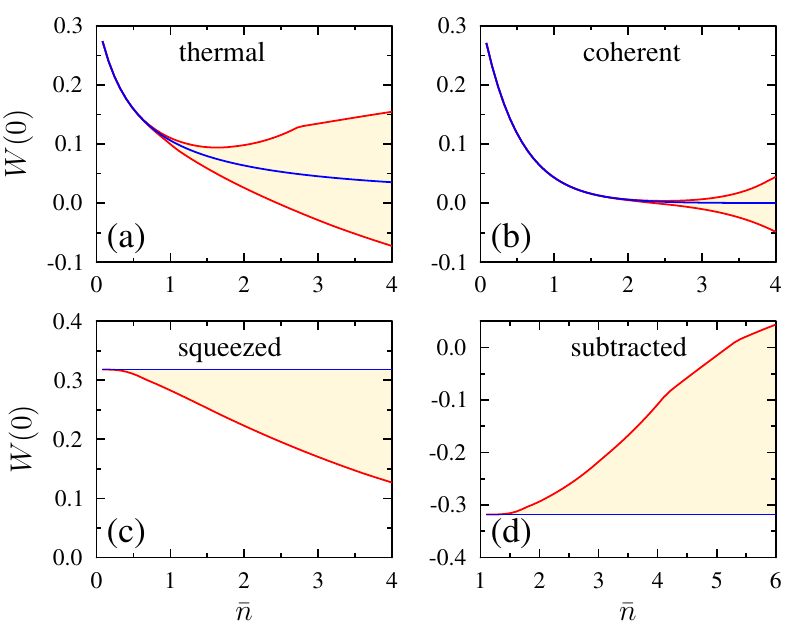}
\caption{Lower and upper bounds on the value of Wigner function at the origin of phase space $W$ are plotted in dependence on the mean photon number $\bar{n}$  for four different states:  thermal state (a), coherent state (b), squeezed vacuum state (c), and single-photon subtracted squeezed vacuum state (d). Blue lines indicate the true values of $W$. Parameters of the detector read $M=10$ and $\eta=1$.}
\label{figWigner}
\end{figure}

As discussed above, the linear programming approach can be used to obtain the fundamental uncertainties of determination of mean value of an arbitrary bounded operator $\hat{Z}$ diagonal in Fock basis. As an important example we consider here determination of the mean value of Wigner function at the origin of phase space,
\begin{equation}
W= \frac{1}{\pi} \sum_{n=0}^\infty (-1)^n p_n.
\end{equation}
\tcr{Note that the value of the Wigner function at the origin of phase space is proportional to the mean value of the photon number parity operator $(-1)^{\hat{n}}$. }
Figure~\ref{figWigner} shows the dependence of the numerically calculated lower and upper bounds $W_{\mathrm{min}}$ and $W_{\mathrm{max}}$ on the mean photon number $\bar{n}$ for four different types of states and fixed number of detection channels $M=10$. Similarly to the behavior of bounds on $p_n$ (c.f. Fig.~\ref{figp5bounds}), the width of the uncertainty window increases with increasing $\bar{n}$. For the photon-subtracted squeezed vacuum state, the considered detection scheme with ten detection channels enables unambiguous certification of negativity of Wigner function up to $\bar{n}\approx 5.15$.
For higher mean photon numbers, the click statistics becomes compatible also with states that have positive Wigner function at the origin of phase space. The squeezed vacuum state and the single-photon subtracted squeezed vacuum state are states with well defined parity of photon number distribution. Therefore, the upper (or lower) bound on $W$ coincides with the true value of $W$ in those two cases.

\begin{figure}[t]
\includegraphics[width=0.8\linewidth]{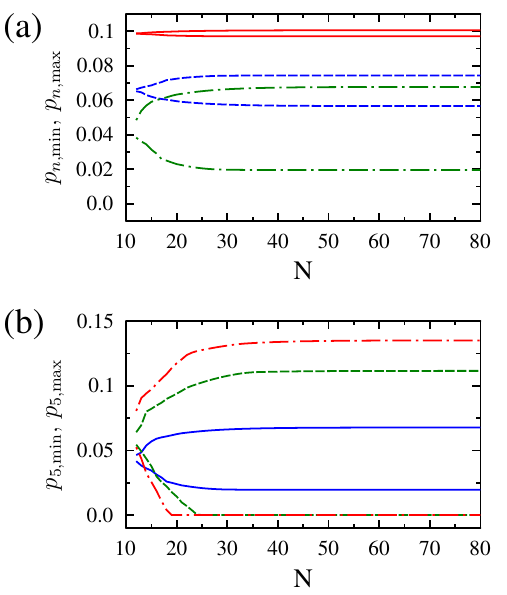}
\caption{Lower and upper bounds on the photon number probabilities  $p_{n,\mathrm{min}}$ and $p_{n,\mathrm{max}}$ of thermal states are plotted in dependence on the photon number cutoff $N$. Parameters of the multiplexed detector read $M=10$ and $\eta=1$. In panel (a) we show results for fixed $\bar{n}=2$  and $n=3$ (red solid line), $n=4$ (blue dashed line) and $n=5$ (green dot-dashed  line). The panel (b) illustrates dependence of the probability bounds on the mean photon number, with bounds on $p_5$ plotted for $\bar{n}=2$ (blue solid line), $\bar{n}=3$ (green dashed line), and $\bar{n}=4$ (red dot-dashed line). }
\end{figure}

\tcr{The value of the Wigner function at some point of phase space $(x,p)$ different from the origin can be detetermined by measuring the mean value of displaced parity operator \cite{Banaszek1996,Banaszek1999}, where the required  coherent displacement reads $\alpha=(1/\sqrt{2})(x+ip)$. The coherent displacement can be implemented for instance by mixing the input signal with an auxiliary coherent beam at a strongly unbalanced beam splitter \cite{Banaszek1996,Banaszek1999}. If the mean value of the photon-number parity operator is estimated from measurements with the multiplexed detector, then one has to take into account that the coherent displacement may inccrease the width of the photon number distribution and may affect and reduce the precision of estimation of the Wigner function value. The data plotted in Fig.~\ref{figWigner}(b) can be equivalently interpreted as probing the value of the Wigner function of vacuum state  at the phase-space circle with diameter $|\alpha|=\sqrt{\bar{n}}$. The graph thus illustrates that for large enough displacement the uncertainty of estimation  of the Wigner function value increases.}

The dependence of the numerically calculated lower and upper bounds $p_{n,\mathrm{min}}$ and $p_{n,\mathrm{max}}$ on the photon number cutoff $N$ is illustrated in Fig.~9. Initially, the uncertainty window $\Delta p_n$ increases with increasing $N$ because more and more degrees of freedom become available for optimization. For large $N$ the numerically calculated values saturate and converge to their true values. For the  range of state parameters considered in the present work the choice $N=80$ turns out to be sufficient to obtain reliable results. Similar convergence behavior is observed also for other bounded quantities, such $W$. However, this picture changes when unbounded operators are considered and we discuss this in the following section.

\section{Estimation of mean photon number}
\begin{figure}[b]
\includegraphics[width=\linewidth]{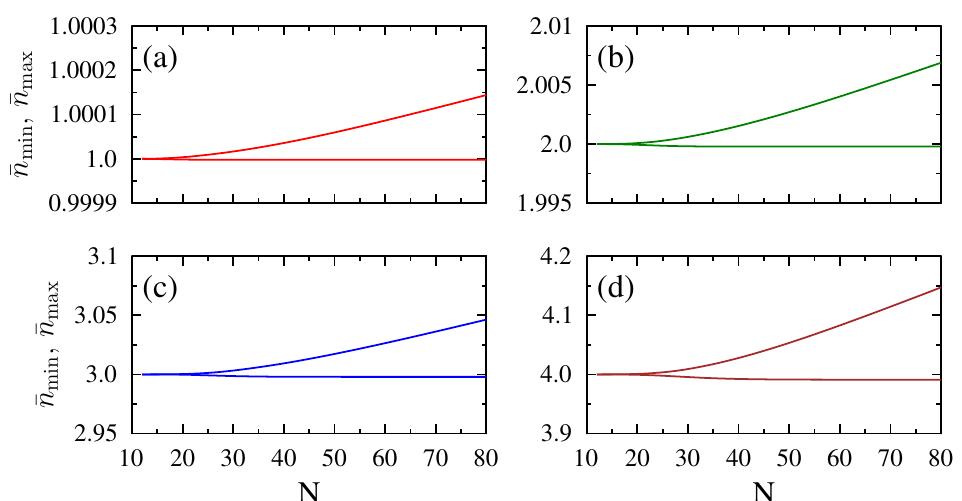}
\caption{The lower and upper bounds on the mean photon number $\bar{n}_{\mathrm{min}}$ and $\bar{n}_{\mathrm{max}}$ are plotted in dependence on the photon number cutoff $N$. The results are shown for thermal state with true mean photon numbers $\bar{n}=1$ (a),  $\bar{n}=2$ (b), $\bar{n}=3$ (c), and  $\bar{n}=4$ (d). Parameters of the multiplexed detector are $M=10$ and $\eta=1$. }
\label{fignbarbound}
\end{figure}

The photon number operator $\hat{n}$ is unbounded. Therefore, without making some additional assumptions, the knowledge of the click statistics $c_M$ for finite $M$ does not bound the mean photon number $\bar{n}$ from above. The reason is that arbitrarily small population of sufficiently large Fock state can increase the mean photon number arbitrarily. For example, if for some fixed $K$ we have $p_{K^2}=1/K$, then this contributes to the mean photon number by the value $K^2 p_{K^2}=K$ that can be arbitrarily large for arbitrarily small $p_{K^2}$.   This effect is illustrated in Fig.~\ref{fignbarbound}, where we plot the dependence of the lower and upper bounds on $\bar{n}$ on the photon number cutoff $N$. We can see that the upper bound increases with increasing $N$ and does not saturate.

Nevertheless, the bound on the mean photon number can be rather tight if we can reliably assume that the contribution of Fock states above some threshold $N$ is negligible. This could follow for instance from the actual physical mechanism of the state generation.  For example, for thermal state with $\bar{n}=2$ and detector with $M=10$ channels we find that the determination uncertainty $\Delta \bar{n} <0.01$ even for the cutoff $N=80$. It is worth noting that the uncertainty of determination of Fock state probabilities is much larger [cf. Fig.~3(a)]. For instance $\Delta p_6 >0.06 $ in this case. However, the possible changes of $p_n$ with respect to their true values are mutually anticorrelated, since $p_n$ are constrained by the values of $c_m$ and by the overall probability normalization. Therefore, the uncertainty of $\bar{n}$  is much smaller for the cutoff $N=80$ than one could naively expect from the uncertainties of $p_n$.

\begin{figure}[t]
\includegraphics[width=\linewidth]{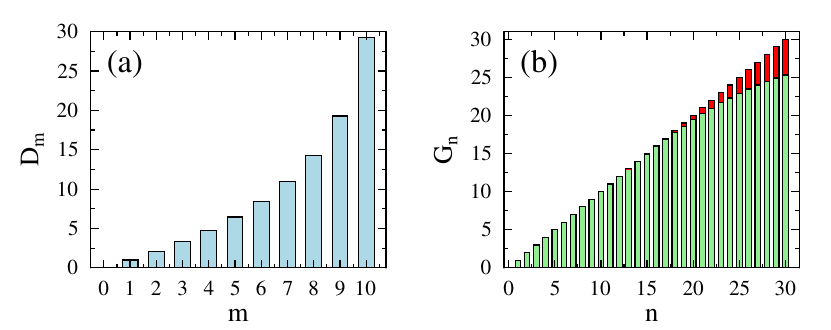}
\caption{Example of estimation of the mean photon number $\bar{n}$ from click statistics, $M=10$, $\eta=1$. Panel (a) shows  a simple estimator $D$ that is exact for states that do not contain more than ten  photons. In panel (b), this estimator is plotted in the photon number basis (green bars), and the difference between the actual values of the coefficients $G_n$  and the ideal values $n$ is indicated by red color.}
\label{fignbarestimator}
\end{figure}

To further understand this, we can consider a simple estimation of the mean photon number
\begin{equation}
\bar{n}_{\mathrm{est}}=\sum_{m=0}^M D_m c_m,
\label{nbarestimator}
\end{equation}
 which  is exact for states containing no more than $M$ photons. The coefficients $D_m$ can be determined by solving system of linear equations
\begin{equation}
\sum_{m=0}^M D_m C_{mn} = n, \qquad n\leq M.
\end{equation}
The coefficients $D_m$ are plotted in Fig.~\ref{fignbarestimator}(a) for detector with $M=10$ channels.  The estimator (\ref{nbarestimator}) can also  be  rewritten as 
\begin{equation}
\bar{n}_{\mathrm{est}}= \sum_{n=0}^\infty G_n p_n,
\end{equation}
where $G_n=n$ for $n\leq M$ and $G_n<n$ for $n>M$. Nevertheless, we find that the performance of the estimator is very good even for several photon numbers above $M$  [see Fig.~\ref{fignbarestimator}(b)].
For example, $G_{15}=14.95$. Asymptotically, $G_n$ saturate and $\lim_{n\rightarrow \infty} G_n=D_M$.

\section{Analytical bounds on single-photon probabilities}

In this section we provide examples of analytical bounds on single-mode  single-photon probability $p_1$ and the two-mode probability $p_{1,1}$ that a single photon is present in each mode.
These bounds are obtained for simple Hanbury Brown-Twiss setups \cite{Kimble1977,Grangier1986} where two on/off detectors are used to measure each mode. The two-mode probability $p_{1,1}$ can  characterize the probability of sucessful generation of correlated pair of photons hence its determination is of high practical relevance. Results  presented and discussed in this section complement the general numerical analysis performed in the previous sections.  Consider first the single-mode case, where the measured probabilities (\ref{q0kdefinition}) read $p_0$ and
\begin{equation}
q_0=\sum_{n=0}^\infty \frac{1}{2^n} p_n.
\label{q0definition}
\end{equation}
The resulting linear constraints on probabilities $p_n$ read
\begin{equation}
\sum_{n=1}^\infty p_n=1-p_0, \qquad \sum_{n=1}^\infty\frac{1}{2^n} p_n=q_0-p_0.
\label{p1constraints}
 \end{equation}
A lower bound on  the single-photon probability $p_1$ based on knowledge of $p_0$ and $q_0$ can be expressed as follows \cite{Straka2011,Hlousek2021}:
\begin{equation}
p_{1,\mathrm{min}}=\max\left(4 q_0-3p_0-1,0\right).
\label{p1minanalytical}
\end{equation}
Let us first assume that $4q_0-3p_0-1\ge0$. The bound (\ref{p1minanalytical}) is saturated by distribution $\tilde{p}_0=p_0,$ $\tilde{p}_1=4q_0-3p_0-1$, $\tilde{p}_2=2(1+p_0-2q_0)$, and $\tilde{p}_n=0$, $n>2$. Using the explicit definition of $q_0$, Eq. (\ref{q0definition}), it is easy to show that $\tilde{p}_1 \leq p_1$ and $\tilde p_2 \geq 0$. Explicitly,
\begin{equation}
\tilde{p}_1=p_1-\sum_{n=3}^\infty\left (1-\frac{4}{2^n}\right)p_n,\qquad \tilde{p}_2=2\sum_{n=2}^\infty\left(1-\frac{2}{2^n}\right) p_n.
\label{ptildeseries}
\end{equation}
The lower bound can be equivalently rewrtiten as 
\begin{equation}
p_{1,\mathrm{min}}=y_1(1-p_0)+y_{2} (q_0-p_0),
\label{p1minz}
\end{equation}
where  $y_1=-1$ and $y_2=4$. These coefficients satisfy the inequalities
\begin{equation}
y_{1}+\frac{1}{2^n}y_{2} \leq \delta_{n1}, \qquad n\geq 1.
\end{equation}
Therefore, $y_1$ and $y_{2}$ represent solution of a linear program that is dual to the minimization of $p_{1}$ under the constraints (\ref{p1constraints}). Moreover,  a feasible solution of the original linear program exists that saturates the bound (\ref{p1minz}), as we have just demonstrated above.  By duality, this implies the optimality of $p_{1,\mathrm{min}}$.  Observe that the lower bound $p_{1,\mathrm{min}}=4q_0-3p_0-1$ is tight for all states that do not contain more than two photons, i.e. $p_n=0$ for $n>2$.

If $4q_0-3p_0-1<0$, then a probability distribution that staturates the bound (\ref{p1minanalytical}) can be constructed as follows: $\tilde{p}_0=p_0$, $\tilde{p}_1=0,$ $\tilde{p}_2=4(q_0-p_0)$, $p_n=0$ for $n>2$  but finite, and  $\tilde{p}_\infty=1+3p_0-4q_0$. This distribution should be understood as the limit of the sequence of distributions with $\tilde{p}_n=0$, $n>2$, $n\neq N$, and $\tilde{p}_N=1+3p_0-4q_0$, when $N\rightarrow \infty$. Note that $\tilde{p}_2 \geq 0$ by construction and also all the constraints (\ref{p1constraints}) are satisfied by construction.

\begin{figure}[t]
\includegraphics[width=0.85\linewidth]{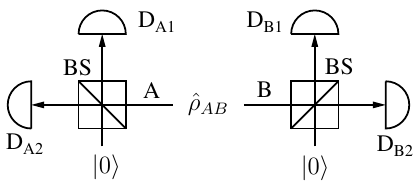}
\caption{Hanbury Brown-Twiss type scheme for characterization of single-mode and two-mode states. Each signal mode is split into two halves at a balanced beam splitter BS and each output mode is measured with a binary detector $D$ that can distinguish the presence and absence of photons. Probabilities of various combinations of clicks or non-clicks of the detectors are measured.}
\label{figHBTsetup}
\end{figure}

We can also construct an upper bound on $p_1$,
\begin{equation}
p_{1,\mathrm{max}}=2(q_0-p_0)=\sum_{n=1}^\infty \frac{2}{2^n} p_n.
\label{p1maxanalytical}
\end{equation}
Therefore, $p_{1,\mathrm{max}}\geq p_1$ by construction.
The upper  bound (\ref{p1maxanalytical}) is saturated by the following photon number distribution $\tilde{p}_0=p_0,$ $\tilde{p}_{1}=2(q_0-p_0)$, $\tilde{p}_{\infty}=1+p_0-2q_0$, and all other $\tilde{p}_{n}$ vanish. It follows from Eq. (\ref{ptildeseries})  that  $\tilde{p}_\infty \geq 0$ for any true photon number distribution $p_{n}$. This proves the optimality of the upper bound (\ref{p1maxanalytical}).

Let us now consider characterization of the joint photon number distribution $p_{m,n}$ of a two-mode state $\hat{\rho}_{AB}$ by local measurements on modes A and B, see Fig.~\ref{figHBTsetup}. We assume that each signal is locally split on a balanced beam splitter and each output mode is measured with an on/off detector. By considering only local responses of the detectors observing mode A (or B), we can measure the following probabilities
\begin{equation}
p_{0A}=\sum_{n=0}^\infty p_{0,n}, \qquad p_{0B}=\sum_{m=0}^\infty p_{m,0},
\label{p0Ap0Bdefinition}
\end{equation}
and 
\begin{equation}
q_{A}=\sum_{m,n=0}^\infty  \frac{1}{2^m} p_{m,n}, \qquad q_{B}=\sum_{m,n=0}^\infty  \frac{1}{2^n} p_{m,n}.
\end{equation}

If we consider also the various correlations between non-clicks of detectors that observe modes A and B, we can measure the two-mode vacuum probability $p_{0,0}$ as well as the probabilities
\begin{equation}
q_{0A}=\sum_{m=0}^\infty \frac{1}{2^{m}} p_{0,m}, \qquad q_{0B}=\sum_{n=0}^\infty \frac{1}{2^{n}}  p_{n,0},
\end{equation}
and
\begin{equation}
q_{00}=\sum_{m,n=0}^\infty \frac{1}{2^{m+n}}\, p_{m,n}.
\label{q00definition}
\end{equation}

To simplify the subsequent analysis, we will restrict ourselves to symmetric states with $p_{m,n}=p_{n,m}$ and we will seek bound on $p_{1,1}$ that reflects this symmetry. 
We can get rid of all the probabilities $p_{m,n}$ where $m=0$ or $n=0$ by constructing the following linear combinations of $p_{n,m}$,
\begin{equation}
D_1=1-p_{0A}-p_{0B}+p_{0,0}=\sum_{m,n=1}^\infty p_{m,n},
\end{equation}
\begin{equation}
D_2=q_{00}-q_{0A}-q_{0B}+p_{0,0},
\label{D2definition}
\end{equation}
and
\begin{equation}
D_3=q_A+q_B-q_{0A}-q_{0B}-p_{0A}-p_{0B}+2p_{0,0}.
\label{D3definition}
\end{equation}
We can express \tcr{$D_2$ and}  $D_3$ as
\begin{equation}
\tcr{D_2= \sum_{m,n=1}^\infty \frac{p_{m,n}}{2^{m+n}},} \qquad D_3= \sum_{m,n=1}^\infty \left(\frac{1}{2^m}+\frac{1}{2^n}\right)  p_{m,n}.
\end{equation}

Based on $D_{k}$, a lower bound on $p_{1,1}$ can be constructed as follows,
\begin{equation}
p_{1,1,\mathrm{min}}=12D_2-2D_3,
\label{p11min}
\end{equation}
which is applicable when $12D_2-2D_3 \geq 0$.  \tcr{We emphasize  that this bound is expressed solely in terms of the click statistics, or equivalently the probabilities of no-clicks, which can be measured with the two-mode setup depicted in Fig.~\ref{figHBTsetup}. See also the definitions of parameters $D_2$ and $D_3$ in Eqs. (\ref{D2definition}) and (\ref{D3definition}).}

A probability distribution that saturates \tcr{the bound (\ref{p11min})} reads $\tilde{p}_{1,1}=12D_2-2D_3$, $\tilde{p}_{1,2}=\tilde{p}_{2,1}=2D_3-8 D_2$, 
$\tilde{p}_{\infty,\infty}=D_1+4D_2-2D_3$, and all other  $\tilde{p}_{m,n}$ with $m,n\geq 1$ vanish. The meaning of $\tilde{p}_{\infty,\infty}$ here is the same as the meaning of $\tilde{p}_{\infty}$ above. Using the explicit expressions for $D_k$ it is not difficult to check that $\tilde{p}_{1,2}\geq 0$
and $\tilde{p}_{\infty,\infty}\geq 0 $ for any true distribution $p_{m,n}$.  Moreover, we have
\begin{equation}
p_{1,1,\mathrm{min}}=\sum_{m,n=1}^\infty \left(\frac{12}{2^{m+n}}-\frac{2}{2^m}-\frac{2}{2^n}\right)p_{m,n} \leq p_{1,1}.
\label{p11mininequality}
\end{equation}
The inequality in Eq. (\ref{p11mininequality}) holds because all the coefficients in front of $p_{mn}$ in the summation are negative or vanish, except for the coefficient in front of $p_{1,1}$, which is equal to $1$.
Since the inequality (\ref{p11mininequality}) can be saturated if $p_{1,1,\mathrm{min}} \geq 0$, this proves the optimality of the bound (\ref{p11min}) based on knowledge of $D_k$.  We stress that we have derived the bound assuming symmetry of the photon number distribution. For general states, full information contained in the  probabilities (\ref{p0Ap0Bdefinition}) --- (\ref{q00definition}) must be utilized to obtain the  optimal bound.

\section{Discussion and Conclusions}

In summary, we have investigated the fundamental bounds on the ability to determine photon number distribution and other related quantities from tomographically incomplete measurements performed with an array of $ M$ binary detectors that can only distinguish the presence and absence of photons. The measured click statistics generally restricts the values of $p_n$ to some interval and we have calculated the lower and upper bounds on $p_n$  by solving appropriate linear programs. We have verified on concrete examples that the precision of determination of the photon number statistics increases with increasing number of detection channels $M$ and decreases with decreasing detection efficiency $\eta$. We have also investigated the determination of parity of the photon number distribution and the estimation of the mean photon number.  In our analysis, we have not made any additional assumptions about the underlying photon number distribution. The proposed approach can be straightforwardly extended to characterization of photon number distributions of multimode fields, although the number of parameters that enter the linear program will rapidly increase with the number of modes.

In our present study, we have deliberately assumed that the true values of the click statistics $\vec{c}=\{c_{m}\}$  are known, which allowed us to fully focus on the effects of tomographic incompleteness. 
 In practice, any measurement will be affected by noise due to finite number of samples. The directly measured values $c_m$  thus can generally differ from the true values and may even be incompatible with any photon number distribution $p_n$.  Applying statistical approaches, we may try to access the posterior distribution of the physically allowed values of $c_m$, determined by the measurement data.  
\tcr{Suppose that we can sample the posterior distribution of $\vec{c}$.   
For each sample $\vec{c}_s$ we can find the lower and upper bounds $z_{\mathrm{min},s}$ and $z_{\mathrm{max},s}$ on the quantity of interest $Z$ by solving the linear program (\ref{linearprogram}). After processing $S$ samples, ensembles of the lower and upper bounds will be established, which will characterize the  uncertainty of determination of $Z$.}
 
 The linear programming approach utilized in this work is applicable to any measurement scheme where the measured probabilities depend linearly on the photon number probabilities and the measurement is tomographically incomplete hence it does not fully determine the distribution $p_n$. The matrix coefficients $C_{mn}$ that connect the photon number statistics $p_n$ and the measured distribution $c_m$ are often determined by some theoretical model, but they can also be obtained by quantum measurement tomography \cite{Luis1999,Fiurasek2001,Lundeen2009,DAuria2011,Brida2012,Brida2012b,Natarajan2013,Cooper2014,Piacentini2015,Ma2016,Izumi2020,Endo2021}. Clearly, any uncertainty in the description of the measurement device will further diminish our ability to infer the photon number statistics from the measured data \tcr{\cite{Rossi2004}}.

\tcr{The method reported in the present paper is not primarily intended to be used for reconstruction of photon number distributions from experimental data but rather for evaluation of the potential performance and limits of a chosen detection scheme. If some experiment targets preparation and characterization of a specific state with photon number distribution $p_n$, then the present method allows one to analyze the influence of the number of detection channels $M$ on the uncertainty of determination of $p_n$ and one can determine how many detectors are required for reliable characterization  of $p_n$. If the measurement device characterizes some unknown state and a photon number distribution $p_{n,\mathrm{exp}}$ is obtained by some reconstruction method from the experimental data, then one can evaluate the bounds $p_{n,\mathrm{min}}$ and $p_{n,\mathrm{max}}$  assuming that $p_{n,\mathrm{exp}}$  is the true distribution and using a theoretical model of the utilized measurement device.  }

The uncertainties of determination of $p_n$ can be suppressed by using sufficiently large number of detection channels $M$. Since  increasing the number of detectors may be costly and undesirable, one can break the symmetry of the setup and introduce unbalanced splitting of the signal among the $M$ detectors. For  instance, one can use a sequence of $M-1$ balanced detectors in the setup in Fig.~1 to achieve effective transmittance $T_m=1/2^k$ for $k$-th detector, with $T_{M-1}=T_{M}$ \cite{Fiurasek2024}.  The total number of detection channels then increases exponentially with $M$  because $2^M-1$ different transmittances can be probed by measuring the probabilities of non-clicks of each combination of the detectors. However, increasing the number of detection channels will result in increased statistical uncertainty of estimation of each particular click or non-click probability. It may be interesting to investigate the trade-off between the effect of  statistical errors and the tomographic incompleteness, as the effective number of  detection channels changes. However, this is beyond the scope of the present paper and we leave it to future work.

\begin{acknowledgments}
The author acknowledges support by the project OP JAC  CZ.02.01.01/00/23\_021/0008790  of the Ministry of Education, Youth, and Sports of the Czech Republic and EU.
\end{acknowledgments}

\section*{Data availability statement}
The data that support the findings of this article are openly available \cite{MultiplexZenodo2025}.


\begin{thebibliography}{10}

\bibitem{Lita2008} 
A. E. Lita, A. J. Miller, and S. W. Nam, Counting near-infrared single-photons with 95\% efficiency, Opt. Express \textbf{16}, 3032 (2008).

\bibitem{Kim1999}
\tcr{J. Kim, S. Takeuchi, Y. Yamamoto, and H. H. Hogue,  Multiphoton detection using visible light photon counter, Appl. Phys. Lett. \textbf{74}, 902 (1999).}

\bibitem{Gerrits2012}
T. Gerrits, B. Calkins, N. Tomlin, A. E. Lita, A. Migdall, R. Mirin, and S. Woo Nam,  Extending single-photon optimized superconducting transition edge sensors beyond the single-photon counting regime, Opt. Express \textbf{20}, 23798 (2012).

\bibitem{Harder2016}
G. Harder, T. J. Bartley, A. E. Lita, S. Woo Nam, T. Gerrits, and C. Silberhorn, Single-Mode Parametric-Down-Conversion States with 50 Photons as a Source for Mesoscopic Quantum Optics,
Phys. Rev. Lett. \textbf{116}, 143601 (2016).



\bibitem{Sperling2017}
J. Sperling, W. R. Clements, A. Eckstein, M. Moore, J. J. Renema, W. S. Kolthammer, S. W. Nam, A. Lita, T. Gerrits, W. Vogel, G. S. Agarwal, and I. A.Walmsley, Detector-Independent Verification of Quantum Light, Phys. Rev. Lett. \textbf{118}, 163602 (2017).

\bibitem{Cahall2017}
C. Cahall, K. L. Nicolich, N. T. Islam, G. P. Lafyatis, A. J. Miller, D. J. Gauthier, and J. Kim, Multi-photon detection using a conventional superconducting nanowire single-photon detector, Optica \textbf{4}, 1534 (2017). 

\bibitem{Magana2019}
 O. S. Magaña-Loaiza, Roberto de J. León-Montiel, Armando Perez-Leija, Alfred B. U’Ren, Chenglong You, Kurt Busch, Adriana E. Lita, S. W. Nam, R. P. Mirin, and  T. Gerrits, 
Multiphoton quantum-state engineering using conditional measurements, Npj Quantum Inf. \textbf{5}, 80 (2019).

\bibitem{Liao2020}
T. Liao, Z. Li, and B. Wang, Direct measurement of the PDC photon statistics by PNR detector, Opt. Commun. \textbf{477}, 126352 (2020). 

\bibitem{Zhu2020}
D. Zhu, M. Colangelo, C. Chen, B. A. Korzh, F. N. C. Wong, M. D. Shaw, and K. K. Berggren, Resolving Photon Numbers Using a Superconducting Nanowire with Impedance-Matching Taper, Nano Letters \textbf{20}, 3858 (2020).


\bibitem{Eaton2022} 
M. Eaton, A. Hossameldin, R. J. Birrittella, P. M. Alsing, C. C. Gerry, H. Dong, C. Cuevas, and O. Pfister, Resolution of 100 photons and quantum generation of unbiased random numbers, 
Nat. Photonics \textbf{17}, 106 (2023).

\bibitem{Davis2022}
S. I. Davis, A. Mueller, R. Valivarthi, N. Lauk, L. Narvaez, B. Korzh, A. D. Beyer, O. Cerri, M. Colangelo, K. K. Berggren, M. D. Shaw, S. Xie, N. Sinclair, and M. Spiropulu, Improved Heralded Single-Photon Source with a Photon-Number-Resolving Superconduct-
ing Nanowire Detector, Phys. Rev. Applied \textbf{18}, 64007 (2022).

\bibitem{Los2024}
 J. W. N. Los, M. Sidorova, B. Lopez-Rodriguez, P. Qualm, J. Chang, S. Steinhauer, V. Zwiller, and I. E. Zadeh, High-performance photon number resolving detectors for 850–950 nm wavelength range, APL Photonics \textbf{9},  066101 (2024).

\bibitem{Paul1996}
H. Paul, P. T\"{o}rm\"{a}, T. Kiss, and I. Jex, Photon Chopping: New Way to Measure the Quantum State of Light, Phys. Rev. Lett. \textbf{76}, 2464 (1996).


\bibitem{Rehacek2003}
J. \v{R}eh\'{a}\v{c}ek, Z. Hradil, O. Haderka, J. Pe\v{r}ina, and M. Hamar, Multiple-photon resolving fiber-loop detector, Phys. Rev. A \textbf{67}, 061801(R)  (2003).

\bibitem{Banaszek2003} 
K. Banaszek and I. A. Walmsley, Photon counting with a loop detector, Opt. Lett. \textbf{28}, 52-54  (2003).

\bibitem{Fitch2003}
M. J. Fitch, B. C. Jacobs, T. B. Pittman, and J. D. Franson, Photon-number resolution using time-multiplexed single-photon detectors, Phys. Rev. A \textbf{68}, 043814 (2003).

\bibitem{Achilles2003}
D. Achilles, C. Silberhorn, C. Sliwa, K. Banaszek, and I. A. Walmsley, Fiber-assisted detection with photon number resolution, Opt. Lett. \textbf{28},  2387 (2003).

\bibitem{Kalashnikov2011}
D. A. Kalashnikov, S. H. Tan, M. Chekhova, and L. A. Krivitsky, Accessing photon bunching with a photon number resolving multi-pixel detector, Opt. Express \textbf{19}, 9352 (2011).


\bibitem{Sperling2012b}
J. Sperling, W. Vogel, and G. S. Agarwal, Sub-Binomial Light, Phys. Rev. Lett. \textbf{109}, 093601 (2012).


\bibitem{Sperling2012a}
J. Sperling, W. Vogel,  and G. S. Agarwal, True photocounting statistics of multiple on-off detectors, Phys. Rev. A \textbf{85}, 023820 (2012).


\bibitem{Bartley2013}
T. J. Bartley, G. Donati, X.-M. Jin, A. Datta, M. Barbieri, and I. A. Walmsley, Direct observation of subbinomial light, Phys. Rev. Lett \textbf{110}, 173602  (2013).


\bibitem{Mattioli2016}
F. Mattioli, Z. Zhou, A. Gaggero, R. Gaudio, R. Leoni, and A. Fiore, Photon-counting and analog operation of a 24-pixel photon number
resolving detector based on superconducting nanowires, Opt. Express  \textbf{24}, 9067 (2016).


\bibitem{Kroger2017}
J. Kr\"{o}ger, T. Ahrens, J. Sperling, W. Vogel, H. Stolz, and B. Hage, High intensity click statistics from a $10 \times 10$
avalanche photodiode array, J. Phys. B: At. Mol. Opt. Phys. \textbf{50}, 214003 (2017).

\bibitem{Zhu2018}
D. Zhu, Q.-Y. Zhao, H. Choi, T.-J. Lu, A. E. Dane, D. Englund, and K. K. Berggren, A scalable multi-photon coincidence detector based on superconducting nanowires, Nat. Nanotechnol. \textbf{13}, 596 (2018).

\bibitem{Straka2018}
I. Straka, L. Lachman, J. Hlou\v{s}ek, M. Mikov\'{a}, M. Mi\v{c}uda, M. Je\v{z}ek and R. Filip, Quantum non-Gaussian multiphoton light, Npj Quantum Inf. \textbf{4}, 4 (2018).



\bibitem{Kovalenko2018}
O. P. Kovalenko, J. Sperling, W. Vogel, and A. A. Semenov, Geometrical picture of photocounting measurements, Phys. Rev. A \textbf{97}, 023845  (2018).

\bibitem{Jonsson2019}
M. J\"{o}nsson and G. Bj\"{o}rk, Evaluating the performance of photon-number-resolving detectors, Phys. Rev. A \textbf{99}, 043822 (2019).


\bibitem{Tiedau2019}
J. Tiedau, E. Meyer-Scott, T. Nitsche, S. Barkhofen, T. J. Bartley, and C. Silberhorn, A high dynamic range optical detector for measuring single photons and bright light, Opt. Express \textbf{27}, 1 (2019).

\bibitem{Lachman2019}
 L. Lachman, I. Straka, J. Hlou\v{s}ek, M. Je\v{z}ek, and R. Filip, Faithful Hierarchy of Genuine n-Photon Quantum Non-Gaussian Light, Phys. Rev. Lett. \textbf{123}, 043601 (2019).



\bibitem{Sperling2017b}
J. Sperling, A. Eckstein, W. R. Clements, M. Moore, J. J. Renema, W. S. Kolthammer, S. W. Nam, A. Lita,
T. Gerrits, I. A. Walmsley, G. S. Agarwal, and W. Vogel, Identification of nonclassical properties of light with multiplexing layouts, Phys. Rev. A \textbf{96}, 013804 (2017).

\bibitem{Hlousek2019}
J. Hlou\v{s}ek, M. Dudka, I. Straka, and M. Je\v{z}ek, Accurate detection of arbitrary photon statistics, Phys. Rev. Lett \textbf{123},  153604 (2019).


\bibitem{Cheng2022}
R. Cheng, Y. Zhou, S. Wang, M. Shen, T. Taher, and H. X. Tang, A 100-pixel photon-number-resolving detector unveiling photon statistics, Nat. Photonics \textbf{17}, 112 (2023).

\bibitem{Knoll2023}
\tcr{L. T. Knoll, G. Petrini, F. Piacentini, P. Traina, S. V. Polyakov, E. Moreva, I. P. Degiovanni, and M. Genovese, 
Photon Statistics Modal Reconstruction by Detected and Undetected Light, Ad. Quantum Technol. \textbf{6}, 2300062 (2023).}



\bibitem{Hlousek2024}
J. Hloušek, J. Grygar, M. Dudka, and M. Ježek, High-resolution coincidence counting system for large-scale photonics applications, Phys. Rev. Applied \textbf{21}, 024023 (2024).


\bibitem{Krishnaswamy2024}
S. Krishnaswamy, F. Schlue, L. Ares, V. Dyachuk, M. Stefszky, B. Brecht, C. Silberhorn, and J. Sperling, Experimental retrieval of photon statistics from click detection, Phys. Rev. A \textbf{110}, 023717  (2024).

\bibitem{Sullivan2024}
N. M. Sullivan, B. Braverman, J. Upham and R. W. Boyd, Photon number resolving detection with a single-photon detector and adaptive storage loop,
New J. Phys. \textbf{26},  043026 (2024).


\bibitem{Santana2024}
T. S. Santana, C. D. Muñoz, R. A. Starkwood, and C. J. Chunnilall, 
Extending the quantum tomography of a quasi-photon-number-resolving detector,  Opt. Express \textbf{32},  20350 (2024).

\bibitem{Banner2024}
P. R. Banner, D. Kurdak, Y. Li, A. Migdall, J. V. Porto, and S. L. Rolston, Number-state reconstruction with a single single-photon avalanche detector,
Optica Quantum \textbf{2}, 110 (2024).



\bibitem{Mogilevtsev1998}
D. Mogilevtsev, Diagonal element inference by direct detection, Opt. Commun. \textbf{156}, 307 (1998).


\bibitem{Rossi2004}
A. R. Rossi, S. Olivares, and M. G. A. Paris, Photon statistics without counting photons, Phys. Rev. A \textbf{70}, 055801 (2004).

\bibitem{Zambra2006}
G. Zambra and M. G. A. Paris, Reconstruction of photon-number distribution using low-performance photon counters, Phys. Rev. A \textbf{74}, 063830 (2006).

\bibitem{Banaszek1996}
\tcr{K. Banaszek and K. Wódkiewicz, Direct Probing of Quantum Phase Space by Photon Counting, Phys. Rev. Lett. \textbf{76}, 4344  (1996).}

\bibitem{Banaszek1999}
\tcr{K. Banaszek, C. Radzewicz,  K. Wódkiewicz, and J. S. Krasiński, Direct measurement of the Wigner function by photon counting, Phys. Rev. A \textbf{60}, 674 (1999).}




\bibitem{Kimble1977}
H. J. Kimble, M. Dagenais, and L. Mandel, Photon antibunching in resonance fluorescence, Phys. Rev. Lett. \textbf{39}, 691 (1977).

\bibitem{Grangier1986}
P. Grangier, G. Roger, and A. Aspect, Experimental evidence for a photon anticorrelation effect on a beam splitter: A new light on single-photon interferences, Europhys. Lett. \textbf{1}, 173 (1986).



\bibitem{Straka2011}
M. Je\v{z}ek, I. Straka, M. Mi\v{c}uda, M. Du\v{s}ek, J. Fiur\'{a}\v{s}ek, and R. Filip, Experimental Test of the Quantum Non-Gaussian Character of a Heralded Single-Photon State,
Phys. Rev. Lett. \textbf{107}, 213602 (2011).


\bibitem{Hlousek2021}
J. Hlou\v{s}ek, M. Je\v{z}ek, and J. Fiur\,{a}\v{s}ek, Direct Experimental Certification of Quantum Non-Gaussian Character and Wigner Function Negativity of Single-Photon Detectors,
Phys. Rev. Lett. \textbf{126}, 043601  (2021).


\bibitem{Luis1999}
A. Luis and L. L. S\'{a}nchez-Soto, Complete Characterization of Arbitrary Quantum Measurement Processes, Phys. Rev. Lett. \textbf{83}, 3573 (1999).

\bibitem{Fiurasek2001}
    J. Fiur\'{a}\v{s}ek, Maximum-likelihood estimation of quantum measurement, Phys. Rev. A \textbf{64}, 024102 (2001).

\bibitem{Lundeen2009}
    J. S. Lundeen, A. Feito, H. Coldenstrodt-Ronge, K. L. Pregnell, Ch. Silberhorn, T. C. Ralph, J. Eisert, M. B. Plenio, and I. A. Walmsley, Tomography of quantum detectors, Nat. Phys. \textbf{5}, 27 (2009).

\bibitem{DAuria2011}
    V. D’Auria, N. Lee, T. Amri, C. Fabre, and J. Laurat, Quantum Decoherence of Single-Photon Counters, Phys. Rev. Lett. \textbf{107}, 050504 (2011).

\bibitem{Brida2012}
    G. Brida, L. Ciavarella, I. P. Degiovanni, M. Genovese, L. Lolli, M. G. Mingolla, F. Piacentini, M. Rajteri, E. Taralli, and M. G. A. Paris, Quantum characterization of superconducting photon counters, New J. Phys. \textbf{14}, 085001 (2012).

\bibitem{Brida2012b}
\tcr{G. Brida, L. Ciavarella, I. P. Degiovanni, M. Genovese, A. Migdall, M. G. Mingolla, M. G. A. Paris, F. Piacentini, and S. V. Polyakov, 
Ancilla-Assisted Calibration of a Measuring Apparatus, Phys. Rev. Lett. \textbf{108}, 253601 (2012).}

\bibitem{Natarajan2013}
C. M. Natarajan, L. Zhang, H. Coldenstrodt-Ronge, G. Donati, S. N. Dorenbos, V. Zwiller, I. A. Walmsley, and R. H. Hadfield, Quantum detector tomography of a time-multiplexed superconducting nanowire single-photon detector at telecom wavelengths,  Opt. Express \textbf{21}, 893 (2013).


\bibitem{Cooper2014}
    M. Cooper, M. Karpiński, and B. J. Smith, Local mapping of detector response for reliable quantum state estimation, Nat. Commun. \textbf{5}, 4332 (2014).


\bibitem{Piacentini2015}
\tcr{F. Piacentini, M. P. Levi, A. Avella, M. López, S. Kück, S. V. Polyakov, I. P. Degiovanni, G. Brida, and M. Genovese, 
Positive operator-valued measure reconstruction of a beam-splitter tree-based photon-number-resolving detector, Opt. Lett. \textbf{40}, 1548 (2015).}


\bibitem{Ma2016}
    J. Ma, X. Chen, H. Hu, H. Pan, E. Wu, and H. Zeng, Quantum detector tomography of a single-photon frequency upconversion detection system, Opt. Express \textbf{24}, 20973 (2016).

\bibitem{Izumi2020}
    S. Izumi, J. S. Neergaard-Nielsen, and U. L. Andersen, Tomography of a Feedback Measurement with Photon Detection,  Phys. Rev. Lett. \textbf{124}, 070502 (2020).

\bibitem{Endo2021}
M. Endo, T. Sonoyama, M. Matsuyama, F. Okamoto, S. Miki, M. Yabuno, F. China, H. Terai, and A. Furusawa, Quantum detector tomography of a superconducting nanostrip photon-number-resolving detector, Opt. Express \textbf{29}, 11728 (2021).


\bibitem{Fiurasek2024}
J. Fiur\'{a}\v{s}ek, Tight nonclassicality criteria for unbalanced Hanbury Brown–Twiss measurement scheme with click detectors, 
Phys. Rev. A \textbf{109}, 033713 (2024).

\bibitem{MultiplexZenodo2025}
J. Fiur\'{a}\v{s}ek, Data for \emph{Fundamental limits on determination of photon number statistics from measurements with multiplexed on/off detectors}, Zenodo, 2025,  https://doi.org/10.5281/zenodo.15087539.




\end{thebibliography}
\end{document}